\documentclass[journal=jctcce,manuscript=article]{achemso}
\usepackage[english]{babel}
\usepackage{microtype}
\usepackage{amsmath,amsthm,amssymb,amsfonts}
\usepackage{mathtools}
\usepackage{siunitx}
\usepackage{tabularx,graphicx}
\usepackage{algorithm2e}
\usepackage{booktabs}
\usepackage{xr}
\usepackage[breaklinks=true,colorlinks=true,linkcolor=blue,urlcolor=blue,citecolor=blue]{hyperref}

\newcommand{\uHEUR}{u^{\text{HLAT}}}

\newcommand{\uKV}{u^{\text{LAT}}}
\newcommand{\wKV}{w^{\text{LAT}}}

\newcommand{\uBBH}{u^{\text{TOR}}}
\newcommand{\wBBH}{w^{\text{TOR}}}

\DeclareSIUnit\angstrom{\text{\AA}}
\DeclareSIUnit\bar{\text{bar}}

\title{Unwrapping NPT Simulations to Calculate Diffusion Coefficients}

\author{Jakob T\'{o}mas Bullerjahn}
\email{jakob.bullerjahn@biophys.mpg.de}
\affiliation{Department of Theoretical Biophysics, Max Planck Institute of Biophysics, 60438 Frankfurt am Main, Germany}

\author{S\"{o}ren von B\"{u}low}
\affiliation{Structural Biology and NMR Laboratory, Linderstr\o m-Lang Centre for Protein Science, Department of Biology, University of Copenhagen, 2200 Copenhagen, Denmark}

\author{Maziar Heidari}
\affiliation{Department of Theoretical Biophysics, Max Planck Institute of Biophysics, 60438 Frankfurt am Main, Germany}

\author{J\'{e}r\^{o}me H\'{e}nin}
\affiliation{Laboratoire de Biochimie Th\'{e}orique UPR 9080, Institut de Biologie Physico-Chimique, CNRS and Universit\'e Paris-Cit\'e, 75005 Paris, France}

\author{Gerhard~Hummer}
\email{gerhard.hummer@biophys.mpg.de}
\affiliation{Department of Theoretical Biophysics, Max Planck Institute of Biophysics, 60438 Frankfurt am Main, Germany}
\alsoaffiliation{Institute of Biophysics, Goethe University Frankfurt, 60438 Frankfurt am Main, Germany}

\begin{document}

\maketitle

\begin{abstract}
    In molecular dynamics simulations in the NPT ensemble at constant pressure, the size and shape of the periodic simulation box fluctuate with time. For particle images far from the origin, the rescaling of the box by the barostat results in unbounded position displacements.  Special care is thus required when a particle trajectory is unwrapped from a projection into the central box under periodic boundary conditions to a trajectory in full three-dimensional space, e.g., for the calculation of diffusion coefficients. Here, we review and compare different schemes in use for trajectory unwrapping.  We also specify the corresponding rewrapping schemes to put an unwrapped trajectory back into the central box. On this basis, we then identify a scheme for the calculation of meaningful diffusion coefficients, which is a primary application of trajectory unwrapping. In this scheme, the wrapped and unwrapped trajectory are mutually consistent and their statistical properties are preserved.  We conclude with advice on best practice for the consistent unwrapping of constant-pressure simulation trajectories and the calculation of accurate translational diffusion coefficients.  
\end{abstract}

\clearpage

\section{INTRODUCTION}

Molecular dynamics (MD) simulations are performed by numerically solving the classical equations of motion for every particle in a given system.  For systems in condensed phase, such as proteins in water, these simulations are usually conducted in volumes of finite size subject to periodic boundary conditions (PBCs).  In constant-volume simulations, one can think of the periodic system either as a single box in which opposite faces are identified under what are also referred to as toroidal boundary conditions, or as an infinite periodic lattice of replicates of the central simulation box.  In the \emph{toroidal view}, a particle leaving the central simulation box placed at the coordinate origin reenters the box at the opposing face, as it would when moving around on a torus.  In the \emph{lattice view}, each particle corresponds to a collection of infinitely many points on a periodic lattice, whose lattice constants are determined by the box size and shape. The toroidal view naturally leads to so-called wrapped trajectories, where particles at every instance in time are contained within the central box (and positions outside the box do not make mathematical sense).  By contrast, in the lattice view each individual marked point on the lattice representing a particular particle can traverse the full three-dimensional space, resulting in an associated unwrapped trajectory.  For simulation boxes of constant volume in constant-energy (NVE) and constant-temperature (NVT) ensembles, the task of unwrapping a trajectory therefore corresponds to transforming from the toroidal view to the lattice view.  

\begin{figure}[htbp]
\begin{center}
\includegraphics{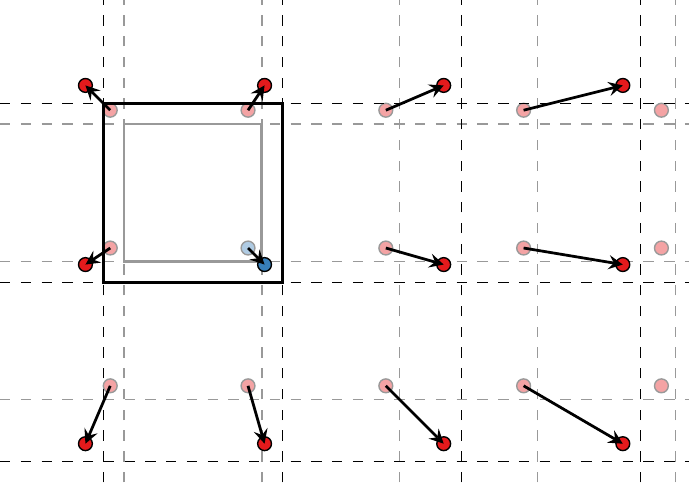}
\caption{Barostat box rescaling in lattice view of PBCs. In the lattice view, the displacement resulting from barostat-induced rescaling of the box volume grows with the distance from the reference box centered at the coordinate origin.  The central boxes before and after barostat action are indicated by gray and black squares, respectively, and the corresponding periodic images of a particle by circles with faint and solid colors.  As a result of barostat rescaling alone, particle images (red) away from the central box move farther than the reference particle (blue).  }
\label{fig:lattice_view}
\end{center}
\end{figure}

In constant-pressure (NPT) simulations, however, the task of unwrapping becomes somewhat ambiguous, because the barostat constantly modifies the size and shape of the simulation box to keep the average pressure fixed. The positions of the particles within the box thereby get rescaled~\cite{Andersen1980}.  In the lattice view of PBCs, the periodic lattice is now fluctuating.  Importantly, the motion of particles purely as a result of the barostat action depends on their distance from the central simulation box and is thus unbounded (see Figure~\ref{fig:lattice_view}).  By contrast, in the toroidal view particles stay in the box with effectively bounded displacements caused by barostat position rescaling.  These differences between the toroidal and lattice views seem to have caused some confusion, as there are at least three different algorithms currently in use to unwrap trajectories of constant-pressure MD simulations.

Here, we review and compare the different schemes proposed for trajectory unwrapping at constant pressure (section~\ref{sec:unwrapping-algorithms}).  We use analytic calculations and numerical examples to demonstrate that lattice-preserving unwrapping schemes give rise to unwrapped trajectories with exaggerated fluctuations when used to unwrap NPT simulation data.  In extreme cases, the dynamics of these unwrapped trajectories differs sharply from the dynamics of the associated wrapped trajectories (sections~\ref{sec:theory} and~\ref{sec:results}).  As a consequence, diffusion coefficient estimates are compromised, an effect that becomes apparent already for bulk water at ambient conditions simulated in the NPT ensemble over a microsecond timescale.  By contrast, we find that a recently proposed off-lattice unwrapping scheme~\cite{von-BulowBullerjahn2020} preserves the statistical properties of the wrapped trajectory and should therefore be preferred for the calculation of translational diffusion coefficients.  However, because the scheme does not adhere to the lattice view, it does not preserve distances~\cite{KulkeVermaas2022}.  Molecules should thus first be made ``whole'' and then unwrapped, e.g., according to their center of mass. We conclude by giving guidance to practitioners on how to extract reliable diffusion coefficient estimates from constant-pressure MD simulations (sections~\ref{sec:best-practice} and~\ref{sec:conlusions}).

\section{UNWRAPPING ALGORITHMS}\label{sec:unwrapping-algorithms}

\paragraph{Heuristic Lattice-View (HLAT) Scheme.}

Some MD simulation and visualization software packages implement a lattice-preserving unwrapping scheme (see, e.g., \texttt{trjconv} in GROMACS~\cite{AbrahamMurtola2015} and \texttt{cpptraj} in Ambertools~\cite{CaseBen-Shalom2018}), which in one dimension (1D) can be cast into the following form:
\begin{equation}\label{eq:heuristic-scheme}
    \uHEUR_{i+1} = w_{i+1} - \left\lfloor \frac{w_{i+1} - \uHEUR_{i}}{L_{i+1}} + \frac{1}{2} \right\rfloor L_{i+1} \, .  
\end{equation}
Here, $w_{i}$ denotes the wrapped position of a particle inside the simulation box of width $L_{i}$ at integration step $i$ corresponding to time $t_i$, $\uHEUR_{i}$ is the corresponding unwrapped position predicted by the HLAT scheme (called the ``heuristic scheme'' in refs~\citenum{von-BulowBullerjahn2020} and~\citenum{KulkeVermaas2022}), and $\lfloor \cdot \rfloor$ denotes the floor function. 
This scheme defines the unwrapped position at time $i+1$ as the particular lattice image of the wrapped position that minimizes the unwrapped displacement from time $i$ to $i+1$, making it intuitively appealing.  In ref~\citenum{von-BulowBullerjahn2020}, however, it was shown that the above scheme occasionally unwraps particles in simulations at constant pressure into the wrong box, which results in an artificial speed up of the particles.  This observation was later confirmed in ref~\citenum{KulkeVermaas2022}.  

\paragraph{Toroidal-View-Preserving ({TOR}) Scheme.} 
After exposing the shortcomings of the HLAT scheme, three of the authors of the present paper proposed an alternative unwrapping scheme, which resolves the issues of eq~\ref{eq:heuristic-scheme} and translates to the following evolution equation in 1D~\cite{von-BulowBullerjahn2020}:
\begin{equation}\label{eq:vonBuelow-Bullerjahn-Hummer-scheme}
    \uBBH_{i+1} = \uBBH_{i} + (w_{i+1} - w_{i}) - \left\lfloor \frac{w_{i+1} - w_{i}}{L_{i+1}} + \frac{1}{2} \right\rfloor L_{i+1} \, .  
\end{equation}
Taking a toroidal view of PBCs, the TOR scheme considers minimal displacement vectors within the simulation box, which are added together to form an unwrapped trajectory.  By design, it therefore preserves the dynamics of the wrapped trajectory.  However, the TOR scheme should only be used to unwrap the trajectories of single particles, such as the center of mass of a molecule or a well-chosen reference atom.  If the scheme is applied separately to multiple atoms of the same molecule, whose intramolecular bonds cross the periodic boundaries, then the atoms in question get incorrectly displaced with respect to each other, resulting in an unphysical stretching of the bonds connecting them together~\cite{KulkeVermaas2022}. Therefore, molecules should first be made ``whole'' and then unwrapped.

\paragraph{Modern Lattice-View (LAT) Scheme.}

An alternative to the HLAT scheme, which takes a lattice view of PBCs without succumbing to the known shortcomings of HLAT, is implemented in the qwrap~\cite{qwrap2016} software package.  To our knowledge, this scheme was never explicitly documented in the literature prior to implementation, but the LAMMPS simulation software~\cite{ThompsonAktulga2022} uses it to write out unwrapped coordinates.  

In the lattice view, crossing the periodic boundaries corresponds to shifting the identity of the particle in the central box to one of its lattice images.  The LAT unwrapping scheme keeps track of these shifts using integer image numbers $n_i$ that indicate how many periodic images the current wrapped coordinates are away from the original, unwrapped particle.  The image number $n_i$ can be obtained either by explicit bookkeeping of image changes due to wrapping (as done by the \texttt{remap} function of LAMMPS), or by detecting large jumps in the wrapped coordinates (as done by the \texttt{qunwrap} feature of qwrap).  In both cases, the unwrapped coordinate can be obtained as a lattice image of its wrapped counterpart, i.e.,
\begin{align}\label{eq:qwrap-scheme}
    \uKV_{i} & = w_{i} - n_{i} L_{i} \, ,
    \\
    \notag
    n_{i+1} & = \sum_{j=0}^i \left\lfloor \frac{w_{j+1}-w_{j}}{L_{j+1}}+\frac{1}{2} \right\rfloor \, .  
\end{align}
This is done in qwrap, and in LAMMPS whenever unwrapped coordinates are necessary, such as for output or for use by the Colvars library~\cite{FiorinKlein2013}.  

Recently, Kulke and Vermaas~\cite{KulkeVermaas2022} proposed a correction to the TOR scheme with the aim to preserve the underlying lattice structure. Their scheme takes the following form in 1D:
\begin{equation}\label{eq:Kulke-Vermaas-scheme}
    \uKV_{i+1} = \uKV_{i} + (w_{i+1} - w_{i}) - \left\lfloor \frac{w_{i+1} - w_{i}}{L_{i+1}} + \frac{1}{2} \right\rfloor L_{i+1} - \left\lfloor \frac{w_{i} - \uKV_{i}}{L_{i}} + \frac{1}{2} \right\rfloor ( L_{i+1} - L_{i} ) \, .  
\end{equation}
However, in hindsight it turns out that eq~\ref{eq:Kulke-Vermaas-scheme} is equivalent to the earlier LAT scheme (eq~\ref{eq:qwrap-scheme}).  This can be seen by substituting eq~\ref{eq:qwrap-scheme} into eq~\ref{eq:Kulke-Vermaas-scheme}, giving
\begin{align*}
    \uKV_{i+1} 
    & = w_{i+1} - n_{i} L_{i} - \left\lfloor \frac{w_{i+1} - w_{i}}{L_{i+1}} + \frac{1}{2} \right\rfloor L_{i+1} - \left\lfloor n_{i} + \frac{1}{2} \right\rfloor ( L_{i+1} - L_{i} )
    \\
    & = w_{i+1} - \left\lfloor \frac{w_{i+1} - w_{i}}{L_{i+1}} + \frac{1}{2} \right\rfloor L_{i+1} - n_{i} L_{i+1} = w_{i+1} - n_{i+1} L_{i+1} \, ,
\end{align*}
where we exploited the relation $\lfloor x + n \rfloor = \lfloor x \rfloor + n$, $n \in \mathbb{Z}$, in the second step.  For this reason, we make no distinction between eqs~\ref{eq:qwrap-scheme} and~\ref{eq:Kulke-Vermaas-scheme}, and refer to them both as the LAT scheme.  

In what follows, we restrict our discussion to the comparison of the TOR and LAT schemes, as the HLAT scheme has already been established as faulty.

\section{THEORY}\label{sec:theory}

Here, we describe a minimal stochastic model of a diffusive particle inside a fluctuating box with PBCs, which we use to generate numerical data and to highlight the differences between the unwrapping schemes via analytic calculations.  We also develop and identify appropriate (re)wrapping schemes for the TOR and LAT schemes, respectively.  

\subsection{Minimal Stochastic Model}

The 1D Gaussian model was introduced in ref~\citenum{von-BulowBullerjahn2020} and provides a minimal theoretical description of constant-pressure MD simulations.  It consists of a Wiener process $w$ that evolves between two periodic boundaries, located at $\pm L_{i}/2$ at time integration step $i$, which are themselves modeled as Gaussian white noise.  Due to box length fluctuations, the value of the process gets rescaled in each time step, after which a diffusive displacement is performed.  The model gives rise to the following wrapped trajectory:
\begin{align}\label{eq:wrapped-trajectory}
    w_{i+1} & = w_i + \bigg( \frac{L_{i+1}}{L_{i}} - 1 \bigg) w_{i} + \sigma_{w} R_{i+1} - \left\lfloor \frac{w_{i}}{L_{i}} + \frac{\sigma_{w} R_{i+1}}{L_{i+1}} + \frac{1}{2} \right\rfloor L_{i+1} \, , 
    \\
    \notag
    L_{i+1} & = \overline{L} + \sigma_{L} S_{i+1} \, ,
\end{align}
where $R_{i}, S_{i} \sim \mathcal{N}(0,1)$ denote uncorrelated normally distributed random variables with zero mean and unit variance, $\overline{L}$ is the average length of the 1D simulation box, and $\sigma_{w}$ and $\sigma_{L}$ determine the noise amplitudes of the random processes driving particle diffusion and box fluctuations, respectively.  

In the absence of wrapping events, the displacements resulting from box rescaling and diffusion would give rise to a trajectory of the following form:
\begin{equation}\label{eq:unwrapped-trajectory}
    u_{i+1} = u_{i} + \bigg( \frac{L_{i+1}}{L_{i}} - 1 \bigg) w_{i} + \sigma_{w} R_{i+1} \, ,
\end{equation}
which can be regarded as the unwrapped partner trajectory to $w$ of eq~\ref{eq:wrapped-trajectory}.  Note that the second term on the right-hand side of eq~\ref{eq:unwrapped-trajectory} represents multiplicative noise, as can best be seen in the limit $\sigma_{L} \ll \overline{L}$, where we have $L_{i+1} / L_{i} - 1 = \sqrt{2} \sigma_{L} S'_{k+1} / \overline{L} + \mathcal{O} ( \sigma_{L}^{2} / \overline{L}^{2} )$ with $S'_{k+1} = (S_{k+1} - S_{k})/\sqrt{2} \sim \mathcal{N}(0,1)$.  Yet, because the noise amplitude is only proportional to $w_{i}$ (and not $u_{i}$), it remains bounded and does not overshadow the diffusive process.

\subsection{Differences Between Unwrapping Schemes}\label{sec:differences-between-schemes}

Unwrapping the wrapped trajectory of eq~\ref{eq:wrapped-trajectory} using the TOR scheme results in an unwrapped trajectory that coincides with eq~\ref{eq:unwrapped-trajectory}.  This can be demonstrated by substituting eq~\ref{eq:wrapped-trajectory} with $n = \lfloor w_{i} / L_{i} + \sigma_{w} R_{i+1} / L_{i+1} + 1 / 2 \rfloor \in \mathbb{Z}$ an integer number into eq~\ref{eq:vonBuelow-Bullerjahn-Hummer-scheme}, giving
\begin{align*}
    \uBBH_{i+1} 
    & = \uBBH_{i} + \bigg( \frac{L_{i+1}}{L_{i}} - 1 \bigg) w_{i} + \sigma_{w} R_{i+1} - \left\lfloor \frac{w_{i}}{L_{i}} - \frac{w_{i}}{L_{i+1}} + \frac{\sigma_{w} R_{i+1}}{L_{i+1}} + \frac{1}{2} \right\rfloor L_{i+1} \, .  
\end{align*}
The remaining floor function evaluates to zero as long as $\sigma_{w} R_{i+1} \ll L_{i+1}$ and $\vert L_{i+1} - L_{i} \vert < L_{i+1} / 2$, which are reasonable assumptions to make for MD simulations when the sampling interval is sufficiently small.  We therefore obtain
\begin{equation*}
    \uBBH_{i+1} \equiv \uBBH_{i} + \bigg( \frac{L_{i+1}}{L_{i}} - 1 \bigg) w_{i} + \sigma_{w} R_{i+1}
\end{equation*}
in all practical cases.  

By contrast, the LAT scheme evaluates to
\begin{equation}\label{eq:almost-unwrapped-trajectory}
    \uKV_{i+1} = \uKV_{i} + \bigg( \frac{L_{i+1}}{L_{i}} - 1 \bigg) w_{i} + \sigma_{w} R_{i+1} - \left\lfloor \frac{w_{i} - \uKV_{i}}{L_{i}} + \frac{1}{2} \right\rfloor ( L_{i+1} - L_{i} )
\end{equation}
when applied to the process of eq~\ref{eq:wrapped-trajectory}.  Here, the last term can be further simplified via eq~\ref{eq:qwrap-scheme}, giving
\begin{align}\label{eq:KV-unwrapped-trajectory}
    \notag
    \uKV_{i+1} & = \uKV_{i} + \bigg( \frac{L_{i+1}}{L_{i}} - 1 \bigg) w_{i} + \sigma_{w} R_{i+1} - \bigg( \frac{L_{i+1}}{L_{i}} - 1 \bigg) n_{i} L_{i}
    \\
    & = \uKV_{i} + \bigg( \frac{L_{i+1}}{L_{i}} - 1 \bigg) \uKV_{i} + \sigma_{w} R_{i+1} \, .  
\end{align}
Comparing eqs~\ref{eq:unwrapped-trajectory} and~\ref{eq:KV-unwrapped-trajectory}, we find that the LAT scheme gives rise to a multiplicative noise term $( {L_{i+1}}/{L_{i}} - 1 ) \uKV_{i}$ that scales with the unwrapped coordinate.  Its magnitude therefore grows without bounds as the particle diffuses away from the origin.  The unbounded multiplicative noise in the LAT scheme causes pathological particle dynamics, which becomes apparent when the LAT scheme is used to unwrap trajectories from Brownian dynamics (BD) and MD simulations, as demonstrated below in section~\ref{sec:results}.

\subsection{Consistent (Re)Wrapping Schemes}\label{sec:wrapping-scheme}

Besides criticizing the undesired effect of intramolecular bond stretching, Kulke and Vermaas~\cite{KulkeVermaas2022} further claimed that the TOR scheme cannot be reversible, because a subsequent wrapping of $\uBBH$ using ``conventional wrapping schemes'' does not reproduce the wrapped trajectory $w$.  While the authors did not explicitly specify which wrapping schemes they are referring to, we expect a lattice-view scheme, which in 1D reads
\begin{equation}\label{eq:KV-wrapping}
    \wKV_{i} = \uKV_{i} - \left\lfloor \frac{\uKV_{i}}{L_{i}} + \frac{\alpha}{2} \right\rfloor L_{i} \, .
\end{equation}
Here, the value of $\alpha$ depends on the definition of the central unit cell.  If it is defined by the interval $[0,L_{i}]$ (as is the case in GROMACS~\cite{AbrahamMurtola2015}) then $\alpha=0$, whereas for cells fluctuating symmetrically around the origin, i.e., $[-L_{i}/2,L_{i}/2]$, one has $\alpha=1$ (this is the convention that LAMMPS~\cite{ThompsonAktulga2022} and NAMD~\cite{PhillipsHardy2020} adhere to).  The scheme in eq~\ref{eq:KV-wrapping} assumes that at each time integration step $i$ the wrapped and unwrapped trajectories are identical up to an integer number of box lengths $L_{i}$, consistent with the lattice view of eq~\ref{eq:qwrap-scheme}.  Equation~\ref{eq:KV-wrapping} should therefore be able to perfectly rewrap a trajectory generated by the LAT unwrapping scheme.  In fact, substituting eq~\ref{eq:almost-unwrapped-trajectory} into eq~\ref{eq:KV-wrapping} with $n = \lfloor (\wKV_{i} - \uKV_{i}) / L_{i} + 1/2 \rfloor$ and $\alpha = 1$ gives
\begin{align*}
    \wKV_{i+1} = \frac{L_{i+1}}{L_{i}} \wKV_{i} + (n - n_{i}) L_{i} + \sigma_{w} R_{i+1} - \left\lfloor \frac{\wKV_{i}}{L_{i}} + (n - n_{i}) \frac{L_{i}}{L_{i+1}} + \frac{\sigma_{w} R_{i+1}}{L_{i+1}} + \frac{1}{2} \right\rfloor L_{i+1} \, ,
\end{align*}
which coincides with eq~\ref{eq:wrapped-trajectory} because
\begin{equation*}
    n = \left\lfloor \frac{\wKV_{i} - \uKV_{i}}{L_{i}} + \frac{1}{2} \right\rfloor = \left\lfloor n_{i} + \frac{1}{2} \right\rfloor \equiv n_{i}
\end{equation*}
must hold.  In light of the fact that the TOR and LAT unwrapping schemes give incompatible results, it is apparent that eq~\ref{eq:KV-wrapping} cannot be used to correctly rewrap $\uBBH$.  

To construct a (re)wrapping scheme consistent with the TOR unwrapping scheme, we backtrace the displacements $\uBBH_{i+1} - \uBBH_{i}$ to reconstruct the wrapped trajectory in an iterative manner as follows:
\begin{equation}\label{eq:vBBH-wrapping}
    \begin{aligned}
    \wBBH_{0} & = \uBBH_{0} \, ,
    \\
    \wBBH_{i+1} & = \wBBH_{i} + (\uBBH_{i+1} - \uBBH_{i}) - \left\lfloor \frac{\wBBH_i + (\uBBH_{i+1} - \uBBH_{i})}{L_{i+1}} + \frac{\alpha}{2} \right\rfloor L_{i+1} \, .  
    \end{aligned}
\end{equation}
Whenever the trajectory $\wBBH_{i} + (\uBBH_{i+1} - \uBBH_{i})$ crosses the periodic boundaries, it gets shifted back into the central box with the help of the last term.  Substituting eq~\ref{eq:unwrapped-trajectory} into eq~\ref{eq:vBBH-wrapping} gives rise to eq~\ref{eq:wrapped-trajectory}, as expected.  Equations~\ref{eq:KV-wrapping} and~\ref{eq:vBBH-wrapping}, and their relations to the LAT and TOR unwrapping schemes, are verified with the help of numerical data in section~\ref{sec:results}.

\section{METHODS}

\subsection{MD Simulation of TIP4P-D Water with GROMACS}\label{sec:gromacs-specifications}

We made use of a \SI{1}{\micro \second} constant-pressure simulation of 515 TIP4P-D water molecules~\cite{PianaDonchev2015} in a cubic box with an average edge length of $\overline{L} \approx \SI{2.5}{\nano\meter}$, which was previously reported on in ref~\citenum{von-BulowBullerjahn2020}. The simulation was run using GROMACS 2018.6~\cite{AbrahamMurtola2015} with a \SI{2}{\femto \second} integration time step, and particle-mesh Ewald electrostatics~\cite{DardenYork1993} with a \SI{1.2}{\nano\meter} real-space cutoff. The SETTLE algorithm was used to keep water molecules rigid~\cite{MiyamotoKollman1992}. The production run commenced after a \SI{100}{\pico \second} initial equilibration at constant volume and a subsequent \SI{5}{\nano\second} equilibration run at constant pressure.  Temperature and pressure were maintained at \SI{300}{\kelvin} and \SI{1}{\bar} throughout the entire simulation using the velocity-rescaling thermostat~\cite{BussiDonadio2007} ($\tau_T=\SI{1}{\pico \second}$) and the Parrinello-Rahman barostat~\cite{ParrinelloRahman1981} ($\tau_p=\SI{5}{\pico \second}$), respectively.  Particle coordinates were recorded every \SI{1}{\pico \second}.

\subsection{MD Simulation of SPC/E Water with LAMMPS}

We generated a set of wrapped and unwrapped trajectories of 511 SPC/E water molecules~\cite{BerendsenGrigera1987} at ambient conditions using the LAMMPS package stable release from 29 September 2021 (update 3)~\cite{Plimpton1995, ThompsonAktulga2022}.  The simulation was performed at constant pressure in a cubic box with an average edge length of $\overline{L} \approx \SI{2.5}{\nano \meter}$.  The SHAKE algorithm~\cite{RyckaertCiccotti1977} was used to constrain the intramolecular bonds and angles at an accuracy tolerance of $10^{-4}$.  The particle-particle particle-mesh solver~\cite{HockneyEastwood1988} with a relative force error accuracy of $10^{-4}$ was used to compute long-range Coulombic interactions, where the cut-off distance in real space was set to \SI{9.8}{\angstrom}.  Equilibration consisted of a \SI{15}{\nano \second} run in the NVT ensemble, followed by a \SI{20}{\nano \second} run in the NPT ensemble. Temperature and pressure were maintained at \SI{300}{\kelvin} and \SI{1}{\bar} using the Nosé-Hoover thermostat and barostat~\cite{Nose1984, Hoover1985} with damping coefficients of 100 and \SI{1000}{\femto \second}, respectively.  The \SI{1}{\micro \second} production run in the NPT ensemble was performed using the same thermostat and barostat coefficients, and a \SI{1}{\femto \second} integration time step.  Particle coordinates of the wrapped and unwrapped trajectory were recorded every \SI{1}{\pico \second} via the \texttt{dump} command.

\subsection{MD Simulation of TIP3P Water with NAMD}

We generated an unwrapped trajectory of 826 water molecules at ambient conditions in a cubic periodic box with $\overline{L} \approx \SI{2.9}{\nano \meter}$, using  NAMD version 3~\cite{PhillipsHardy2020}.  A time step of \SI{2}{\femto \second} was used.  Temperature was maintained at \SI{300}{\kelvin} using underdamped Langevin dynamics with a damping time of \SI{1}{\pico \second}.  Pressure was set to \SI{1}{\bar} using the Nosé-Hoover Langevin piston method as implemented in NAMD~\cite{PhillipsBraun2005}, with a piston period of \SI{200}{\femto \second} and a decay time of \SI{100}{\femto \second}.  Water molecules were kept rigid using the SETTLE algorithm~\cite{MiyamotoKollman1992}.  Long-range electrostatic interactions were computed using the Particle-Mesh Ewald method, with a \SI{12}{\angstrom} cutoff for the real-space part.  The same cutoff was applied to Lennard-Jones potentials, with force-switching for a continuous decay of the force to zero.  Particle coordinates were recorded every \SI{1}{\pico \second} for \SI{900}{\nano \second} in total, this slightly shorter duration being the result of numerical instability (see further section~\ref{sec:lammps-namd-simulations}).  In order to obtain a corresponding wrapped trajectory, we chose to wrap the NAMD output trajectory using eq~\ref{eq:KV-wrapping}.

\section{RESULTS AND DISCUSSION}\label{sec:results}

\subsection{Brownian Dynamics Simulations}\label{sec:bd-simulations}

\begin{figure}[t!]
\begin{center}
\includegraphics{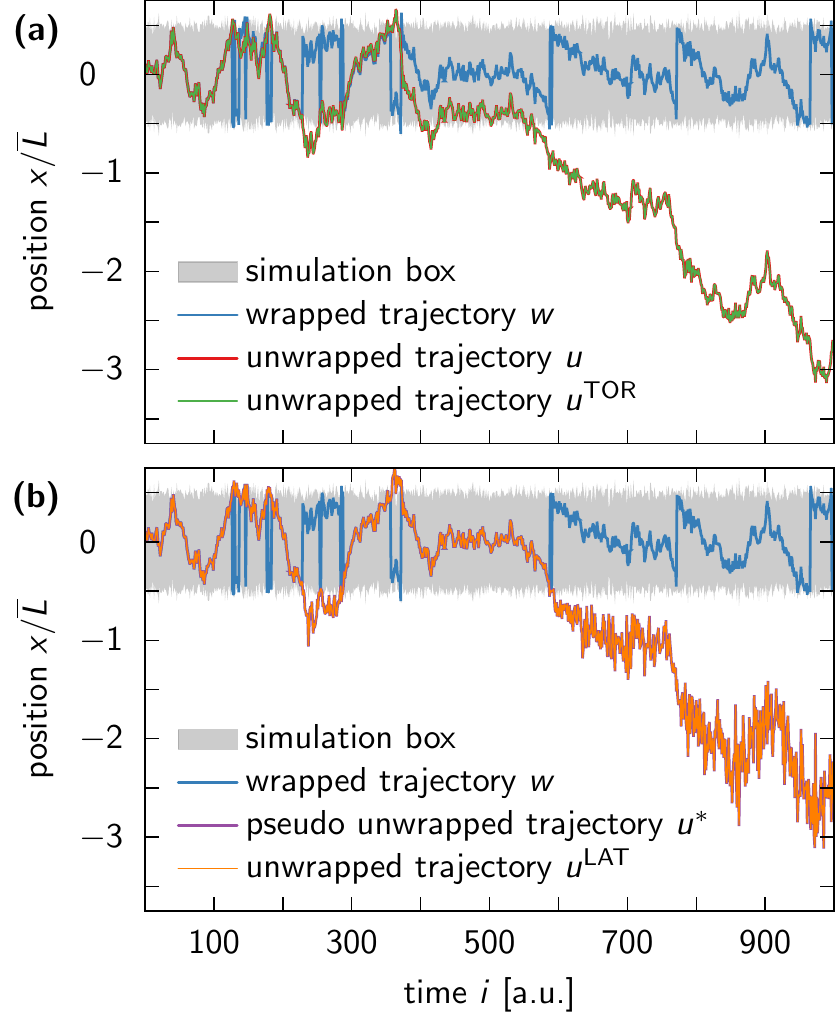}
\caption{Comparison of the TOR and LAT unwrapping schemes in 1D.  (a)~TOR unwrapping of trajectory for 1D Gaussian model.  While the wrapped trajectory $w$ (blue line, eq~\ref{eq:wrapped-trajectory}) is confined to the simulation box (gray shaded area), its unwrapped partner trajectory $u$ (red line, eq~\ref{eq:unwrapped-trajectory}) can traverse arbitrarily far from their common initial position.  The TOR unwrapping scheme (eq~\ref{eq:vonBuelow-Bullerjahn-Hummer-scheme}), when applied to $w$, produces a trajectory $\uBBH$ (green line), which completely overlaps with $u$.  Note that the unwrapped trajectories $u$ and $u^{\text{TOR}}$ are not ``on lattice'' in NPT simulations.  As a result, they may not coincide with the wrapped trajectory $w$ in re-visits to the central simulation box, as seen around time 500.  (b)~LAT unwrapping of the same trajectory as in (a). The unwrapped trajectory $\uKV$ (orange line) generated by the LAT unwrapping scheme (eq~\ref{eq:Kulke-Vermaas-scheme}) coincides with the pseudo unwrapped trajectory $u^{*}$ (purple line, eq~\ref{eq:KV-unwrapped-trajectory}) and exhibits the same exaggerated fluctuations away from the central box.  }
\label{fig:1D_gaussian_model_unwrapping}
\end{center}
\end{figure}

\begin{figure}[t!]
\begin{center}
\includegraphics{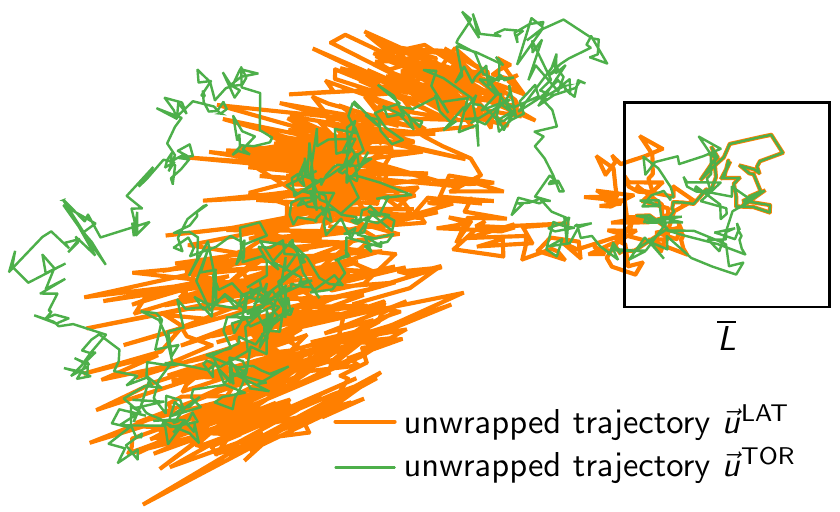}
\caption{Comparison of the TOR and LAT unwrapping schemes in 2D.  A wrapped trajectory $\vec{w}$ of the Gaussian model (not shown) was unwrapped using the TOR and LAT schemes, which resulted in the unwrapped trajectories $\vec{u}^{\text{TOR}}$ (green line) and $\vec{u}^{\text{{LAT}}}$ (orange line), respectively.  While $\vec{u}^{\text{TOR}}$ is visually indistinguishable from an ordinary diffusive trajectory anywhere in the plane, $\vec{u}^{\text{{LAT}}}$ is strongly affected by box fluctuations after leaving the central simulation box (average size shown as black square).  Note that the apparent noise in $\vec{u}^{\text{{LAT}}}$ grows with the distance from the central box and becomes anisotropic, with position fluctuations emanating in a star-like fashion from the origin.  }
\label{fig:2D_gaussian_model}
\end{center}
\end{figure}

\begin{figure}[t!]
\begin{center}
\includegraphics{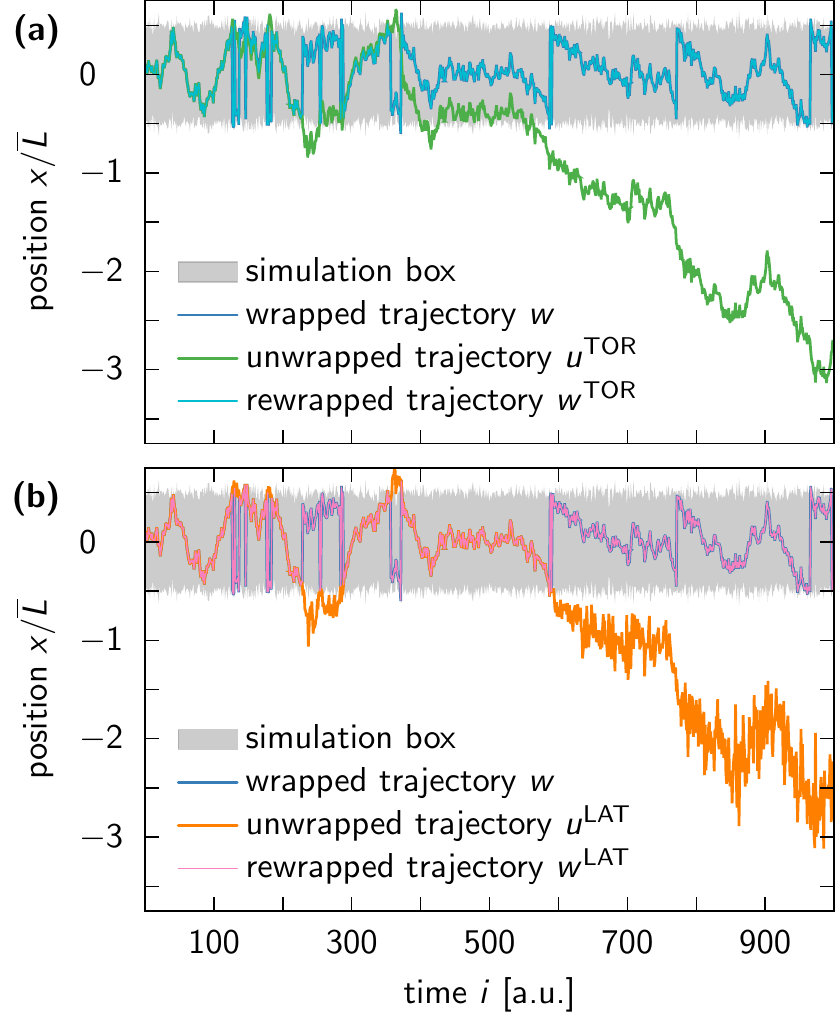}
\caption{Rewrapping trajectories using appropriate wrapping schemes.  (a)~The unwrapped trajectory $\uBBH$ (green line), generated by the TOR scheme, can be perfectly rewrapped inside the simulation box (gray shaded area) using eq~\ref{eq:vBBH-wrapping} with $\alpha = 1$, as seen by the complete overlap of $\wBBH$ (cyan line) with the original wrapped trajectory $w$ (blue line).  (b)~Similar results can be achieved for $\uKV$ (orange line), generated by the LAT scheme, if it is rewrapped using eq~\ref{eq:KV-wrapping} with $\alpha = 1$.  This gives rise to the trajectory $\wKV$ (pink line).  }
\label{fig:1D_gaussian_model_wrapping}
\end{center}
\end{figure}

To verify the analytic predictions of section~\ref{sec:theory}, we evaluated eqs~\ref{eq:wrapped-trajectory}, \ref{eq:unwrapped-trajectory}, and~\ref{eq:KV-unwrapped-trajectory} in an iterative fashion to generate a wrapped trajectory $w$, and two unwrapped trajectories $u$ and $u^{*}$.  The wrapped trajectory was unwrapped using the TOR and LAT schemes (eqs~\ref{eq:vonBuelow-Bullerjahn-Hummer-scheme} and~\ref{eq:Kulke-Vermaas-scheme}, respectively), and the resulting unwrapped trajectories were compared to the corresponding $u$ and $u^{*}$ realizations.  We used random initial positions $w_{0} = u_{0} = u_{0}^{*} = \overline{L} R'$ with $R' \sim \mathcal{U}_{[0,1]}$ uniformly distributed on the interval $[0,1]$, and fixed parameter values of $\sigma_{L} = 0.1 \overline{L}$ and $\sigma_{w} = 0.05 \overline{L}$.  

Figure~\ref{fig:1D_gaussian_model_unwrapping} displays a representative set of trajectories that were generated as described above.  In accordance to our predictions in section~\ref{sec:differences-between-schemes}, the unwrapped trajectory $\uKV$ associated with the LAT scheme exhibits the same position-dependent fluctuations that can be found in $u^{*}$ (eq~\ref{eq:KV-unwrapped-trajectory}), which increase with the distance to the origin in stark contrast to the dynamics of the wrapped trajectory.  Meanwhile, the unwrapped trajectory $\uBBH$ generated by the TOR scheme shows moderate fluctuations and completely overlaps with $u$, as expected.  A visual comparison of trajectory segments between two boundary-crossing events demonstrates that $\uBBH$ perfectly captures the trends observed in $w$.  The same cannot be said about $\uKV$.  

The nondiffusivity of $\uKV$ is even more pronounced in higher dimensions, as illustrated in Figure~\ref{fig:2D_gaussian_model}, where we combine two 1D trajectories of the Gaussian model, $w_{x}$ and $w_{y}$, to construct a two-dimensional (2D) wrapped trajectory $\vec{w} = (w_{x}, w_{y})^{T}$.  The unwrapped trajectories $\vec{u}^{\text{TOR}}$ and $\vec{u}^{\text{LAT}}$ were generated by unwrapping each component of $\vec{w}$ separately using the TOR and LAT schemes, respectively.  As shown, the apparent noise in the $\vec{u}^{\text{LAT}}$ trajectory not only grows with distance from the central simulation box but also becomes anisotropic.  Supporting Movie S1 visualizes the evolution of the trajectories and the fluctuations of the simulation box.

Finally, we assessed the ability of eqs~\ref{eq:KV-wrapping} and~\ref{eq:vBBH-wrapping} to reverse the operations of the LAT and TOR unwrapping schemes, respectively.  In Figure~\ref{fig:1D_gaussian_model_wrapping}, we demonstrate that eq~\ref{eq:vBBH-wrapping} faithfully reproduces the wrapped 1D trajectory $w$ when applied to $\uBBH$.  Similarly, we find that eq~\ref{eq:KV-wrapping} perfectly rewraps $\uKV$ back into the simulation box.  It is therefore unsurprising that Kulke and Vermaas~\cite{KulkeVermaas2022} only found their own unwrapping scheme to be reversible with respect to ``conventional wrapping schemes'' (such as eq~\ref{eq:KV-wrapping}): the unwrapped trajectories $\uBBH$ and $\uKV$ are different and therefore require different wrapping schemes to be correctly rewrapped into the simulation box.

\subsection{GROMACS Simulations}\label{sec:gromacs-simulations}

\begin{figure}[t!]
\begin{center}
\includegraphics{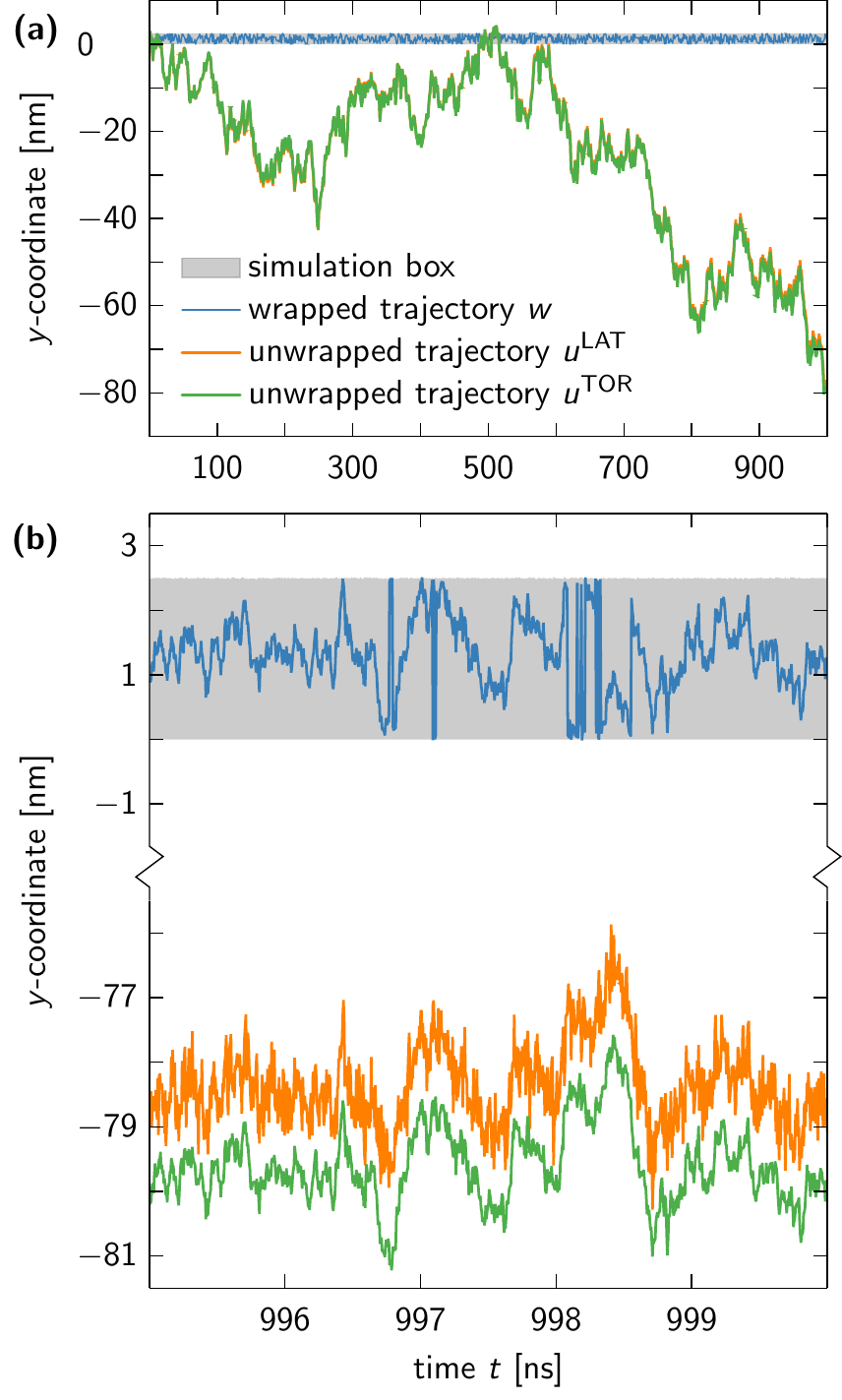}
\caption{Trajectory of an oxygen atom of a TIP4P-D water molecule along a single coordinate axis.  (a)~The wrapped trajectory $w$ (blue line) is unwrapped via the TOR and LAT schemes, resulting in $\uBBH$ (green line) and $\uKV$ (orange line), respectively.  The two unwrapped trajectories seem almost identical, because box fluctuations in MD simulations of water at ambient conditions are small compared to the dimensions of the simulation box (gray shaded area).  (b)~However, a zoom-in on the last 5 nanoseconds of the trajectory reveals that $\uKV$ exhibits larger fluctuations between subsequent time frames than $\uBBH$ and $w$.  The enhanced noise in $\uKV$ is indicative of the unbounded multiplicative noise associated with the LAT unwrapping scheme.  }
\label{fig:gromacs_simulations}
\end{center}
\end{figure}

\begin{figure}[t!]
\begin{center}
\includegraphics{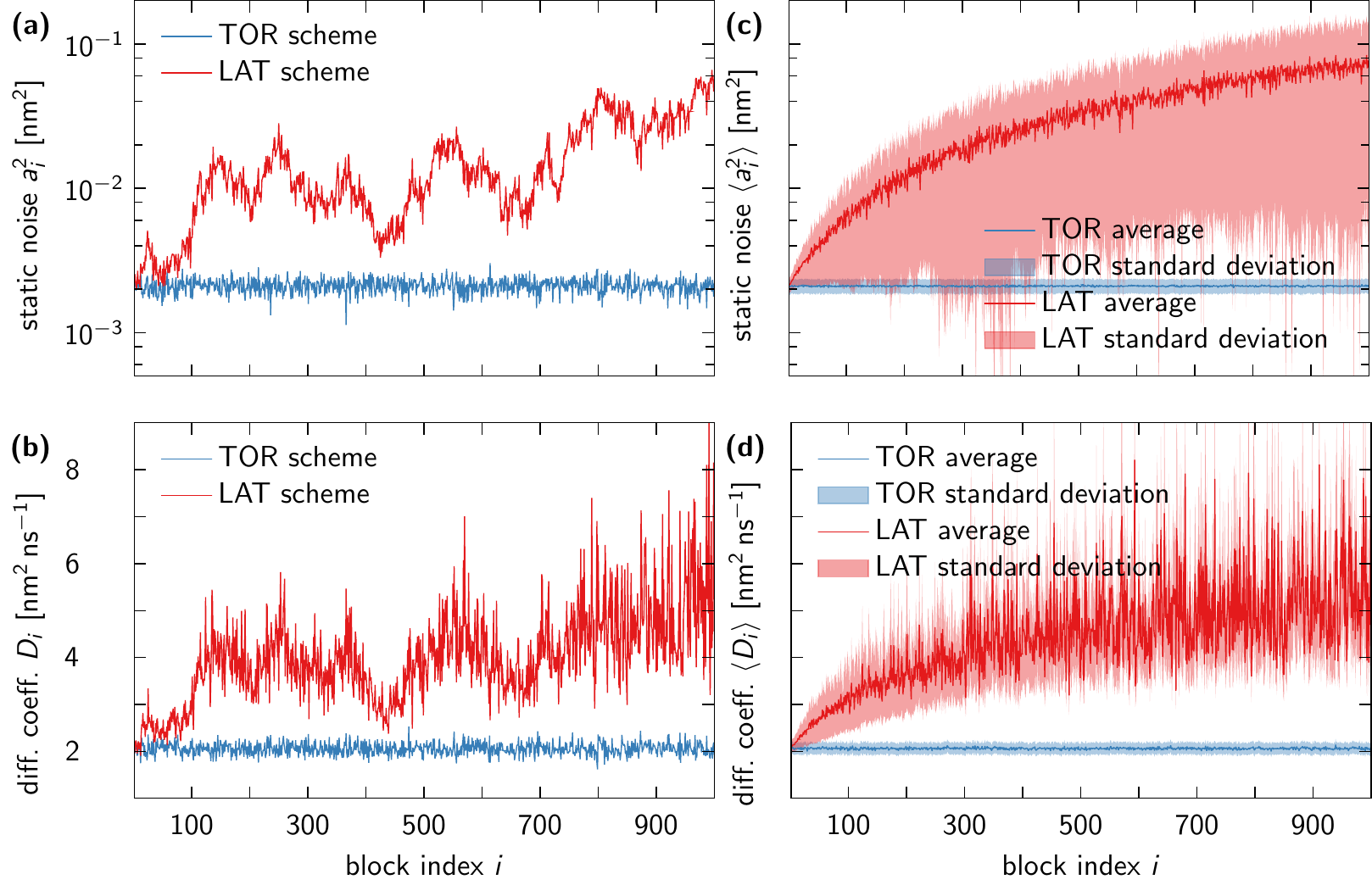}
\caption{Diffusion coefficient estimates are robust for TOR unwrapping, but compromised by LAT unwrapping. Shown are results for the static noise and diffusion coefficient estimates obtained from trajectories of oxygen atoms in TIP4P-D water saved at a time interval of \SI{1}{\pico \second} and divided into 1000 blocks $i$ of \SI{1}{\nano\second} each. (a)~Static noise amplitude $a_{i}$ and (b)~diffusion coefficient $D_{i}$ estimated for each block $i$ of the unwrapped trajectory of a single water molecule using the TOR (blue lines) and LAT schemes (red lines). (c)~Average static noise amplitude $a_{i}$ and (d)~diffusion coefficient $D_{i}$ estimates over all water molecules using the TOR (blue) and LAT schemes (red).  In (a-d), averages are shown as solid lines and standard deviations as shaded areas.  }
\label{fig:diffusion_coefficients}
\end{center}
\end{figure}

In our BD simulations, we could freely choose the amplitude $\sigma_{L}$ of the box fluctuations to highlight the difference between the two unwrapping schemes.  In MD simulations, however, box fluctuations for aqueous systems at ambient conditions are generally well below one percent of the average edge length, so the amplified fluctuations in $\uKV$ are much more subtle.  To test whether we can identify considerable differences between the TOR and LAT schemes in MD simulations, we analyzed the wrapped trajectories of TIP4P-D water in a small cubic box, as reported previously in ref~\citenum{von-BulowBullerjahn2020} (see section~\ref{sec:gromacs-specifications} for technical details).  

In Figure~\ref{fig:gromacs_simulations}, we plot the $y$-component of the trajectory of an oxygen atom in a water molecule that managed to diffuse more than 30 box edge lengths away from the central simulation box.  At first glance, the unwrapped trajectories produced by the TOR and LAT schemes may seem identical, but when we zoom in on the last few nanoseconds of the trajectory, we find $\uKV$ to have the same exaggerated fluctuations as observed in our BD simulations.  By contrast, $\uBBH$ visually reproduces the features of the wrapped trajectory $w$.  

To quantify the effect that box fluctuations have on unwrapped trajectories, we analyzed the diffusive behavior observed in different trajectory segments.  We made use of a maximum likelihood estimator~\cite{BullerjahnHummer2021} (MLE) for the diffusion coefficient $D$, which accounts for the fact that a $d$-dimensional diffusive process $\vec{X}(t)$ can be corrupted by static noise and dynamic motion blur, resulting in the following mean squared displacement (MSD):
\begin{equation*}
    \operatorname{MSD}(\tau) = \langle [\vec{X}(\tau) - \vec{X}(0)]^2 \rangle = a^2 + 2 d D (\tau - 2 B) \, .  
\end{equation*}
Here, $\tau$ denotes the time interval between consecutive saved points along the trajectory.  For MD simulations, the motion blur coefficient $B$ is zero and the vertical intercept $a^2$ accounts for nondiffusive dynamics at short times~\cite{Bullerjahnvon-Bulow2020}.  Note that the MLE does not rely explicitly on estimates for the MSD, but instead exploits the statistics of the increments $\vec{X}(t_{i+1}) - \vec{X}(t_{i})$.  We segmented our unwrapped oxygen trajectories for TIP4P-D water into \SI{1}{\nano \second} blocks and extracted for each block (with index $i$) estimates for the static noise $a_{i}^{2}$ and the diffusion coefficient $D_{i}$.  Figure~\ref{fig:diffusion_coefficients} presents our results for the TOR and LAT schemes.  Unsurprisingly, the TOR scheme gives consistent parameter estimates for all block indices $i$, whereas the estimates for the LAT scheme vary strongly with $i$ and tend towards larger values at later times in the trajectory.  This behavior is to be expected as the diffusive spread of the water molecules moves them further away from the central simulation box, where artifacts become more pronounced in trajectories associated with the LAT unwrapping scheme.  

It should be noted that our results imply a significant difference between global diffusion coefficient estimates obtained for the TOR scheme and the LAT scheme with a \SI{1}{\pico \second} time step.  Consistent with the analysis of Figure~\ref{fig:diffusion_coefficients}, we find the global mean estimates $\overline{D}_{\text{TOR}} = \SI{2.0602 \pm 0.0002}{\nano\meter \squared \per \nano \second}$ and $\overline{D}_{\text{LAT}} = \SI{4.64 \pm 0.03}{\nano\meter \squared \per \nano\second}$ by applying the MLE to the full \SI{1}{\micro\second} trajectory of each oxygen atom and then average over all water molecules, without correcting for finite-size effects~\cite{YehHummer2004, VogeleHummer2016}.  The standard error of $\overline{D}$ was estimated by assuming that the diffusion processes of individual water molecules are uncorrelated.  Importantly, we expect the discrepancy between $\overline{D}_{\text{TOR}}$ and $\overline{D}_{\text{LAT}}$ to grow if the MD simulations are extended beyond the \SI{1}{\micro\second} used here, because $D$-estimates for the LAT unwrapped trajectories will slowly drift to ever larger values as the traced particles move further away from the central box (see Figure~\ref{fig:diffusion_coefficients}).  Our findings contradict the results of ref~\citenum{KulkeVermaas2022}, where no significant difference was found between the two schemes.  One possible explanation for this discrepancy is the fact that Kulke and Vermaas~\cite{KulkeVermaas2022} estimated $D$ from the slope of the MSD without accounting for an intercept $a^{2}$ nor correlations between MSD-values at different times~\cite{Bullerjahnvon-Bulow2020}.  By contrast, the MLE accounts for these subtleties, makes full use of the available data, and thus overall outperforms ordinary least-squares MSD fitting~\cite{BullerjahnHummer2021}.

\subsection{LAMMPS and NAMD Molecular Dynamics Simulations}\label{sec:lammps-namd-simulations}

\begin{figure}[t!]
\begin{center}
\includegraphics{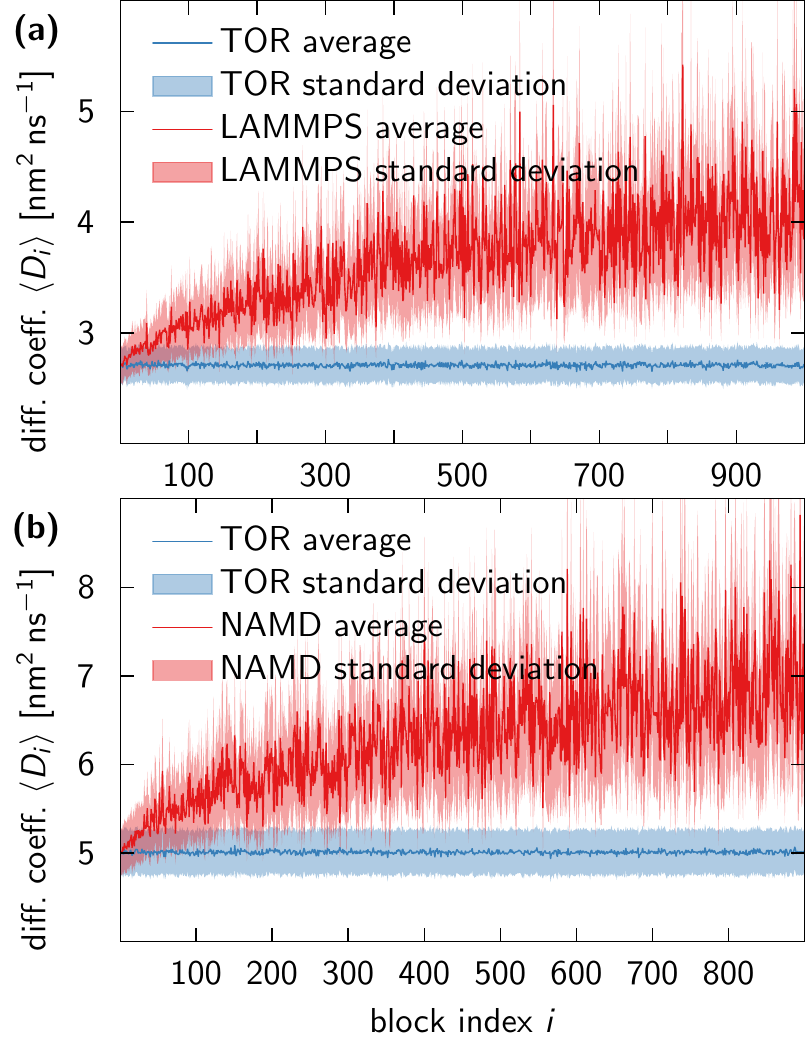}
\caption{Diffusion coefficient estimates for default-unwrapped NAMD trajectories and automatically unwrapped LAMMPS trajectories are compromised. (a)~LAMMPS simulation of SPC/E water.  Diffusion coefficients were estimated separately for 1000 consecutive blocks $i$, each \SI{1}{\nano \second} long, of a continuous trajectory created by writing out ``unwrapped'' coordinates (red).  A corresponding wrapped trajectory, also written out by LAMMPS, was unwrapped using the TOR scheme, and analyzed analogous to the automatically unwrapped LAMMPS trajectory to produce the TOR estimates of the diffusion coefficient (blue).  (b)~NAMD simulation of TIP3P water.  The data analysis procedure was analogous to that used for the LAMMPS trajectories, except that the wrapped counterpart of the NAMD trajectory was generated using eq~\ref{eq:KV-wrapping}.  Due to the fact that the NAMD trajectory was \SI{100}{\nano \second} shorter than the LAMMPS trajectory, it was split into 900 blocks.  
}
\label{fig:lammps_namd_simulations}
\end{center}
\end{figure}

The GROMACS simulation software package exclusively generates wrapped trajectories. These trajectories are typically unwrapped in a post-processing step using built-in tools like \texttt{trjconv} or third-party software, such as the PBCTools~\cite{pbctools2022} and qwrap~\cite{qwrap2016} plugins for VMD~\cite{HumphreyDalke1996}.  Other MD simulation codes write out unwrapped trajectories directly, either by default or via user-specified settings, but this raises the question which unwrapping scheme these trajectories correspond to.  We therefore analyzed simulation trajectories generated via the software packages LAMMPS and NAMD.  NAMD does not, in general, wrap the particle coordinates throughout the simulation, except when writing coordinates to disk, and then only when instructed to do so through the user options \texttt{wrapAll} or \texttt{wrapWater}.  LAMMPS, by contrast, allows the user to specify whether the wrapped coordinates, unwrapped coordinates, or both should be written out.  

We segmented the unwrapped trajectories of oxygen atoms generated by LAMMPS and NAMD into \SI{1}{\nano \second} blocks and analyzed the diffusive dynamics of every block separately, as detailed in section~\ref{sec:gromacs-simulations}.  This was also done to the corresponding wrapped partner trajectories, after unwrapping them via the TOR unwrapping scheme.  The resulting diffusion coefficient estimates as functions of the time window are shown in Figure~\ref{fig:lammps_namd_simulations}.

We find that the diffusion coefficients calculated directly from the unwrapped trajectories steadily diverge to ever larger value as a function of time.  By contrast, the wrapped trajectories, which were unwrapped using the TOR scheme, produce robust diffusion coefficient estimates.  The behavior in Figure~\ref{fig:lammps_namd_simulations} mirrors that seen in Figures~\ref{fig:diffusion_coefficients}c and~\ref{fig:diffusion_coefficients}d, and is fully consistent with LAMMPS and NAMD producing unwrapped NPT trajectories that are ``on lattice'' and thus not suitable for the estimation of diffusion coefficients.  Indeed, the global mean diffusion coefficients obtained by analyzing the entire unwrapped trajectories produced directly (i.e., without division into blocks) by LAMMPS and NAMD are inconsistent with the results for early and late blocks.  The global mean is  $\overline{D}_{\text{LAMMPS}} = \SI{3.75 \pm 0.01}{\nano\meter \squared \per \nano\second}$ for SPC/E water in the LAMMPS simulation, whereas early and late blocks give values around  $\SI{2.7}{\nano\meter \squared \per \nano \second}$ and $\SI{4}{\nano\meter \squared \per \nano \second}$, respectively (see Figure~\ref{fig:lammps_namd_simulations}a). For TIP3P water in the NAMD simulations, the global mean is $\overline{D}_{\text{NAMD}} = \SI{6.50 \pm 0.02}{\nano\meter \squared \per \nano\second}$, with early and late blocks around $\SI{5}{\nano\meter \squared \per \nano \second}$ and $\SI{7}{\nano\meter \squared \per \nano \second}$, respectively (see Figure~\ref{fig:lammps_namd_simulations}b).  By contrast, the global mean $\overline{D}_{\text{TOR}} = \SI{2.7078 \pm 0.0002}{\nano\meter \squared \per \nano \second}$ obtained after TOR unwrapping of the wrapped LAMMPS trajectory is consistent with the respective block estimates, as is $\overline{D}_{\text{TOR}} = \SI{5.0132 \pm 0.0003}{\nano\meter \squared \per \nano \second}$ obtained for the NAMD trajectory.  Note that we did not correct for significant system-size effects on the self-diffusion coefficients~\cite{YehHummer2004, VogeleHummer2016}.  

The NAMD simulation became numerically unstable as particle coordinates and their barostat-induced fluctuations became large.  While such a small box with less than 1000 water molecules can be seen as an extreme example, this phenomenon highlights a benefit of propagating wrapped coordinates internally during the simulation, which is to make the best use of limited floating-point precision, especially in mixed-precision GPU software. In NAMD this behavior can be approached by enabling the \texttt{wrapAll} or \texttt{wrapWater} options, resulting in all coordinates or specifically water molecules being wrapped at every restart.

\subsection{Correctly applying the TOR scheme to bonded atoms}\label{sec:best-practice}

Now that we have established the correctness and consistency of the TOR unwrapping scheme for single particles, we next address possible issues that can arise due to the fact that the scheme does not adhere to the lattice view and therefore cannot preserve distances between particles.  This becomes a problem, for example, when the TOR scheme is applied naively to bonded particles.  

Kulke and Vermaas~\cite{KulkeVermaas2022} observed that bond lengths only got distorted when the TOR scheme was applied to trajectories generated by the GROMACS software package, whereas NAMD trajectories seemed unaffected.  The reason for this discrepancy is the fact that NAMD by default treats molecules as ``whole'' when writing out data, i.e., the software does not break up molecules that sit on the periodic boundary.  A pre-processing step in the analysis of GROMACS data, to make molecules whole prior to unwrapping, will therefore remedy the seeming shortcoming of the TOR scheme observed in ref~\citenum{KulkeVermaas2022}.  

Irrespective of the simulation software behind the data to be analyzed, we recommend the following order of operations when unwrapping MD simulation data of molecules:
\begin{enumerate}
    \item In each frame of the trajectory, make the molecule ``whole,'' i.e., starting from a chosen reference atom of the molecule, ensure that all covalent bonds correspond to their minimal distance over the periodic images. 
    \item Calculate the center of mass of the ``whole'' molecule and, in case the resulting coordinate is located outside of the simulation box, perform a wrapping operation.  This generates a wrapped trajectory of the center-of-mass coordinate of the molecule.
    \item Unwrap the trajectory of the center-of-mass coordinate using the TOR unwrapping scheme.  
    \item If needed, the molecule can be reconstructed along the unwrapped center-of-mass trajectory by using the positions of the atoms relative to the center of mass of the ``whole'' molecule in each frame.  
\end{enumerate}
Note that the calculation of the center-of-mass coordinate in step 2 can be avoided by using instead the position of a specific atom as reference, say the oxygen atom of a water molecule.  Also note that the estimation of translational diffusion coefficients only requires the tracking of the center of mass or any chosen reference atom.

\subsection{Pair diffusion}

\begin{figure}[t!]
\begin{center}
\includegraphics{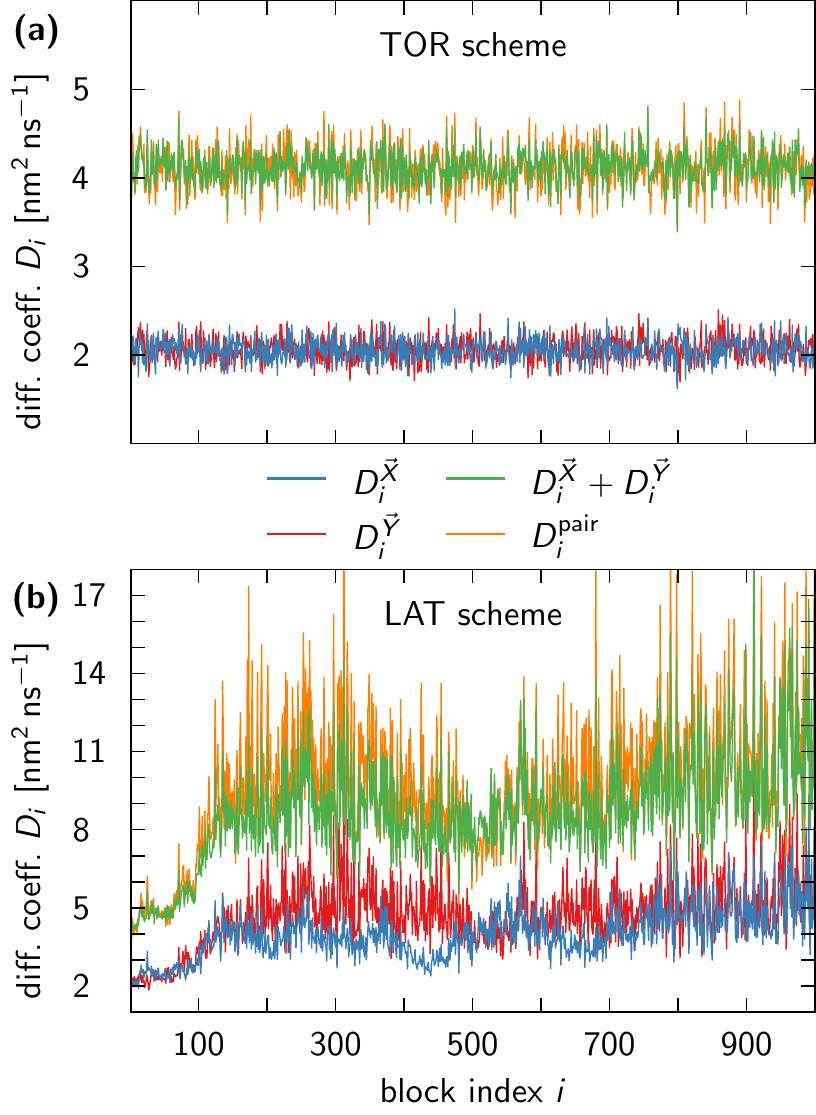}
\caption{Pair diffusion coefficient estimates for the TOR and LAT scheme both satisfy the additivity relation (eq~\ref{eq:pair-diffusion-coefficient}), but are only robust for the TOR scheme.  (a)~Analogous to Figure~\ref{fig:diffusion_coefficients}a, we analyzed \SI{1}{\nano\second} segments of two unwrapped  trajectories generated by the TOR scheme, and extracted the corresponding diffusion coefficients $D_{i}$ (blue and red lines).  The pair diffusion coefficient $\smash{D_{i}^{\text{pair}}}$ (orange line) agrees well with $\smash{D_{i}^{\vec{X}}} + \smash{D_{i}^{\vec{Y}}}$ (green line), but numerical discrepancies are due to the nonlinearity of our MLE.  (b)~Same as in (a) for trajectories unwrapped using the LAT scheme.  }
\label{fig:pair_diffusion_coefficients}
\end{center}
\end{figure}

Finally, one might speculate whether the non-preservation of distances in the TOR unwrapping scheme affects other observables that rely on unwrapping, such as pair diffusion coefficients.  According to theory, the distance vector $\vec{X} - \vec{Y}$ between two independent diffusion processes, $\vec{X}(t)$ and $\vec{Y}(t)$, is also diffusive with the following diffusion coefficient:
\begin{equation}\label{eq:pair-diffusion-coefficient}
    D^{\text{pair}} = D^{\vec{X}} + D^{\vec{Y}} \, .  
\end{equation}
Here, $\smash{D^{\vec{Z}}}$ denotes the diffusion coefficient of the process $\vec{Z}(t)$.  

To test whether pair diffusion is preserved for the TOR scheme, the LAT scheme, or both, we considered two randomly selected TIP4P-D water molecules from the GROMACS MD simulation and analyzed the diffusive behavior of $\vec{X}$, $\vec{Y}$, and $\vec{X} - \vec{Y}$, as described in section~\ref{sec:gromacs-simulations}.  Figure~\ref{fig:pair_diffusion_coefficients} demonstrates that both unwrapping schemes essentially satisfy eq~\ref{eq:pair-diffusion-coefficient}, but only the TOR scheme gives consistent results for all trajectory segments. As for single-particle diffusion, the pair diffusion coefficient obtained by LAT unwrapping tends to grow with time. It is therefore clear that LAT unwrapping and the associated lattice view of the PBCs are not beneficial for the estimate of pair diffusion coefficients in constant-pressure NPT simulations.

\section{CONCLUSIONS}\label{sec:conlusions}

Unwrapping trajectories of constant-pressure MD simulations is a nontrivial task.  One school of thought takes a toroidal view of the PBCs to construct an unwrapped trajectory by adding up minimal displacement vectors at each time step~\cite{von-BulowBullerjahn2020}.  Another~\cite{qwrap2016, KulkeVermaas2022} takes a lattice view of the PBCs and traces the trajectory through the fluctuating lattice of image particles.  As a consequence of the fluctuations in box size and shape, and the associated fluctuations in the lattice parameters, the two approaches produce different unwrapped trajectories.  We have shown here that the toroidal approach embodied in the TOR algorithm~\cite{von-BulowBullerjahn2020} sacrifices the preservation of interparticle distances to preserve the statistical properties of the wrapped trajectory.  In particular, a trajectory created by a diffusion process retains its diffusive character.  By contrast, the lattice view taken in the LAT algorithm~\cite{KulkeVermaas2022} preserves distances between particles, but distorts the statistical properties of local dynamics and thus destroys the diffusive character of a wrapped diffusion trajectory.  

Our analytic calculations (section~\ref{sec:differences-between-schemes}), as well as our results from BD and MD simulations (sections~\ref{sec:bd-simulations} and~\ref{sec:gromacs-simulations}), demonstrate that the lattice-preserving LAT unwrapping scheme amplifies position-dependent fluctuations, which arise in wrapped trajectories of constant-pressure simulations due to barostat position rescaling.  Meanwhile, the TOR unwrapping scheme manages to preserve the diffusive character of the wrapped trajectories.  These observations are further confirmed by our diffusion analysis of unwrapped MD trajectories, where different segments of TOR trajectories give consistent diffusion coefficient estimates, whereas the diffusive dynamics at the beginning and end of LAT trajectories differ greatly (see Figure~\ref{fig:diffusion_coefficients} for water at ambient conditions).  

A surprising conclusion is that the ``unwrapped'' trajectories written out by MD simulation software like NAMD and LAMMPS for NPT simulations should not be used to calculate diffusion coefficients (section~\ref{sec:lammps-namd-simulations}).  The reason is that these trajectories are ``on lattice,'' i.e., the corresponding wrapped positions are obtained by eq~\ref{eq:KV-wrapping}.  Therefore, as the particles diffuse away from the reference box at the origin, they increasingly pick up the multiplicative noise resulting from box rescaling, as visualized in Figures~\ref{fig:1D_gaussian_model_unwrapping} and~\ref{fig:2D_gaussian_model}.  To avoid the resulting artifacts (see, e.g., Figure~\ref{fig:lammps_namd_simulations}) and to obtain accurate diffusion coefficients, the output trajectories should first be wrapped ``on-lattice'' via eq~\ref{eq:KV-wrapping} and then unwrapped ``off-lattice'' using the TOR scheme (eq~\ref{eq:vonBuelow-Bullerjahn-Hummer-scheme}).  

In light of the fact that particle dynamics in constant-pressure simulations is always affected by a (bounded) barostat-induced multiplicative noise, even in the wrapped trajectory, one might be tempted to construct a post-processing scheme to remove this noise altogether.  However, such sanitation is highly nontrivial and its advantage over performing simulations at constant volume is not evident.  For this reason, we instead take a toroidal view of the PBCs and treat the wrapped particle dynamics as the ``true'' dynamics corresponding to the given simulation ensemble, despite the presence of multiplicative noise.  If the multiplicative noise associated with barostat position rescaling is considered an issue, we recommend constant-volume simulations instead of the complex and somewhat arbitrary post-processing of trajectories generated at constant pressure.  

Although current MD simulations of large biological molecules are only minimally affected by the shortcomings of the LAT scheme reported here, we expect our findings to become crucial for the proper analysis of future simulation trajectories on the time scales of milliseconds and beyond.  For NPT simulations, we recommend the use of large boxes, for which the time to diffuse over multiple box dimensions is large and the position rescaling effects are small.  The latter follows from the decay in the relative fluctuations of the box volume $V$ with system size, namely $\smash{\big\langle (V - \langle V \rangle)^2 \big\rangle} \propto \chi_T \langle  V \rangle$, where $\chi_T$ denotes the isothermal compressibility~\cite{Hill1986}.  

For precision calculations of diffusion coefficients and related quantities, one may want to resort to NVT simulations. The choice of an NVT ensemble is advisable in particular for long simulations with small boxes of highly compressible systems with low viscosity, where the relative box-size fluctuations are large and particles can diffuse over many box widths. At constant volume, the lattice and toroidal view of periodic boundary conditions coincide and unwrapping is unambiguous. If needed, the results from NVT simulations at different volumes can be interpolated to the targeted pressure or rigorously combined into weighted samples of an NPT ensemble~\cite{Wood1968, AllenTildesley1987}.  An added advantage of working at constant volume is that the box size and shape entering the large finite-size corrections of translational diffusion coefficients~\cite{YehHummer2004, VogeleHummer2016} are well defined, whereas for NPT conditions one has to resort to averages.

\section*{Supporting Information}

Supporting Video S1 compares the time evolution of 2D trajectories generated via the TOR and LAT schemes.

\begin{acknowledgement}

We thank Dr.~Bal\'{a}zs F\'{a}bi\'{a}n for insightful discussions.  J.T.B., M.H. and G.H. acknowledge financial support by the Max Planck Society.  
J.H. acknowledges support from the French Agence Nationale de la Recherche under grant DYNAMO (ANR-11-LABX-0011) and from Laboratoire International Associ\'e UIUC-CNRS.  S.v.B. acknowledges support from the EMBO Postdoctoral Fellowship ALTF 810-2022.

\end{acknowledgement}

\end{document}